\documentclass[12pt]{article}
\usepackage{graphicx}

\newcommand\pubdate{\today}

\newcommand\pubnumber{LHCb-PROC-2011-043}
\def\Title#1{\begin{center} {\Large #1 } \end{center}}
\def\Author#1{\begin{center}{ \sc #1} \end{center}}
\def\Address#1{\begin{center}{ \it #1} \end{center}}

\newcommand\pubblock{\rightline{\begin{tabular}{l} \pubnumber\\
         \pubdate  \end{tabular}}}
\newenvironment{Abstract}{\begin{center}{\bf Abstract}\end{center} \bigskip \begin{quotation}  }{\end{quotation}}
\newenvironment{Presented}{\begin{quotation} \begin{center} 
             PRESENTED AT\end{center}\bigskip 
      \begin{center}\begin{large}}{\end{large}\end{center} \end{quotation}}
\def\Acknowledgements{\bigskip  \bigskip \begin{center} \begin{large}
             \bf ACKNOWLEDGEMENTS \end{large}\end{center}}

\def\beq{\begin{equation}}
\def\eeq#1{\label{#1}\end{equation}}
\def\eeqn{\end{equation}}

\def\beqa{\begin{eqnarray}}
\def\eeqa#1{\label{#1}\end{eqnarray}}
\def\eeqan{\end{eqnarray}}

\let\bar=\overbar

\def\D{{\cal D}}

\def\Dslash{\not{\hbox{\kern-4pt $D$}}}
\def\dslash{\not{\hbox{\kern-2pt $\del$}}}

\def\BR{\mbox{\rm BR}}

\def\msb{{\bar{\ssstyle M \kern -1pt S}}}

\def\rhobar{\overline{\rho}}
\def\etabar{\overline{\eta}}
\newcommand{\decay}[2]{\ensuremath{#1\!\to #2}}         % {\Pa}{\Pb \Pc}
\def\PK     {\ensuremath{K}}                 
\def\kaon  {\ensuremath{\PK}}
\def\Km    {\ensuremath{\kaon^-}}
\def\PB      {\ensuremath{B}}                 
\def\Bm    {\ensuremath{\PB^-}}
\def\Bp    {\ensuremath{\PB^+}}
\def\Bbar    {\kern 0.18em\overline{\kern -0.18em \PB}{}}
\def\Bsb     {\ensuremath{\Bbar^0_s}}
\def\kaon  {\ensuremath{\PK}}
  \def\Kbar  {\kern 0.2em\overline{\kern -0.2em \PK}{}}

\def\Kp    {\ensuremath{\kaon^+}}
\def\Km    {\ensuremath{\kaon^-}}
\def\KS    {\ensuremath{\kaon^0_{\rm\scriptscriptstyle S}}} 
\def\KL    {\ensuremath{\kaon^0_{\rm\scriptscriptstyle L}}} 
\def\Kstarz  {\ensuremath{\kaon^{*0}}}

\def\PD      {\ensuremath{D}}                 
\def\Dbar    {\kern 0.2em\overline{\kern -0.2em \PD}{}}
\def\D       {\ensuremath{\PD}}

\def\Dz      {\ensuremath{\D^0}}
\def\Dzb     {\ensuremath{\Dbar^0}}
\def\Ds      {\ensuremath{\D^+_s}}
\def\Dstar   {\ensuremath{\D^*}}
\def\Pb     {\ensuremath{b}}                 
\def\b     {\ensuremath{\Pb}}
\def\bbar  {\ensuremath{\overline \b}}
\def\bbbar {\ensuremath{\b\bbar}}
\def\PB     {\ensuremath{B}}                 
\def\B       {\ensuremath{\PB}}
\def\Bs      {\ensuremath{\B^0_s}}
\def\Bdb     {\ensuremath{\Bbar^0}}
\def\Ppi         {\ensuremath{\pi}}                 
\def\pion  {\ensuremath{\Ppi}}
\def\pip   {\ensuremath{\pion^+}}
\def\pim   {\ensuremath{\pion^-}}
\def\Prho      {\ensuremath{\rho}}
\def\rhoz   {\ensuremath{\Prho^0}}
\def\BtoDzK {\decay{B}{\Dz K}}
\def\BtoDzbK {\decay{B}{\Dzb K}}
\def\BmtoDzKm   {\decay{\Bm}{\Dz \Km}}
\def\BsToDsK  {\decay{\Bs}{\Ds K}}
\def\BsbartoDKstar      {\decay{\Bsb}{\Dz \Kstarz}}
\def\BdbartoDRho        {\decay{\Bdb}{\Dz \rhoz}}

\newcommand{\jprlBase}       {Phys.\ Rev.\ Lett.}
\newcommand{\jprBase}       {Phys.\ Rev.}
\newcommand{\jplBase}        {Phys.  Lett. }
\newcommand{\jprd}[1] {\jprBase\ D~{\bf #1}}
\newcommand{\jprl}[1]{\jprlBase\ {\bf #1}}
\newcommand{\plb}[1]{\jplBase  B~{\bf #1}}
\begin{document}
\begin{titlepage}
\pubblock

\vfill

%%%%%%%%%%%%%%%%%%%%%%%%%%%%%%%%%%%%%%%%%%%%%%%%%%%%%%%
%%MODIFY
%%%%%%% TITLE, AUTHOR, ADDRESS 
%%%%%%%%%%%%%%%%%%%%%%%%%%%%%%%%%%%%%%%%%%%%%%%%%%%%%%%

\Title{$\gamma / \phi_3$ at hadron colliders}
\vfill
\Author{Marie-H\'el\`ene Schune (LHCb)  for the  CDF and LHCb collaborations }  
\Address{LAL, Universit\'e Paris-Sud, CNRS/IN2P3, 91898 Orsay, France}
\vfill

%%%%%%%%%%%%%%%%%%%%%%%%%%%%%%%%%%%%%%%%%%%%%%%%%%%%%%%
%%MODIFY
%%%%%%% Abstract
%%%%%%%%%%%%%%%%%%%%%%%%%%%%%%%%%%%%%%%%%%%%%%%%%%%%%%%

\begin{Abstract}
In the first part of the document I  describe in a general way the $\gamma / \phi_3$ extraction and compare the experimental environments.
 I then switch to the available results from the CDF experiment. In the third part I present early results from the LHCb experiment, 
which are promising first steps on the way to a future  $\gamma / \phi_3$ measurement.
\end{Abstract}

\vfill

\begin{Presented}
The Ninth International Conference on\\
Flavor Physics and CP Violation\\
(FPCP 2011)\\
Maale Hachamisha, Israel,  May 23--27, 2011
\end{Presented}
\vfill

\end{titlepage}
\def\thefootnote{\fnsymbol{footnote}}
\setcounter{footnote}{0}
%

%%%%%%%%%%%%%%%%%%%%%%%%%%%%%%%%%%%%%%%%%%%%%%%%%%%%%%%
%%%%%%% Article body
%%%%%%%%%%%%%%%%%%%%%%%%%%%%%%%%%%%%%%%%%%%%%%%%%%%%%%%

\section {Introduction}
Since its first observation in \KL\ decays~\cite{bib:CP_KL}, CPV has been observed and studied in several K and B meson decay 
modes~\cite{bib:PDG}. In the standard model of particle physics (SM)
CP violation in the quark sector of the weak interactions arises from a single irreducible phase
in the Cabibbo-Kobayashi-Maskawa (CKM) matrix \cite{bib:CKM} that describes the mixing of the quarks.
\begin{equation}
V_{\rm{CKM}} =  
\left(
\begin{array}{ccc}
V_{ud} & V_{us} & V_{ub} \\ 
V_{cd} & V_{cs} & V_{cb} \\ 
V_{td} & V_{ts} & V_{tb} \\ 
\end{array}
\right)
\label{eq:CKM}
\end{equation}
There is no theoretical predictions for the different values of the matrix elements which thus need to be determined experimentally. 
In fact the various CKM matrix elements are rather different in size. This strong hierarchy is clearly visible
using the improved Wolfenstein parametrization which is an expansion in power of $\lambda$ : 
\begin{equation}
V_{\rm{CKM}} = 
\left(
\begin{array}{ccc}
1-\frac{\lambda^2}{2}-\frac{\lambda^4}{8}& \lambda &A\lambda^3(\rho -i \eta) \\ 
-\lambda +\frac{A\lambda^5}{2}(1-2\rho)-iA^2\lambda^5 \eta & 1-\frac{\lambda^2}{2}-\lambda^4(\frac{1}{8}+\frac{A^2}{2}) & A\lambda^2 \\
A\lambda^3(1-(1-\frac{\lambda^2}{2})(\rho+i\eta))& -A\lambda^2(1-\frac{\lambda^2}{2})(1+\lambda^2(\rho+i\eta)) & 1 -\frac{A^2\lambda^2}{2}\\
\end{array}
\right) 
 + {\cal O}(\lambda^6)
\label{eq:CKM-rhoeta}
\end{equation}
with $\lambda = s_{12} \simeq 0.22$. 
The $\lambda, A, \rho$ and $\eta$ parameters are known with different accuracies~: $\lambda$ is known to about 0.5~\%, 
$A$ to about 2\%, $\eta$ to about 5\% and $\rho$ only to 15 to 20~\%\footnote{In several cases one uses the variables  
 $\rhobar$ and $\etabar$ $\rhobar = \rho(1-\lambda^2/2)$ and $\etabar = \eta((1-\lambda^2/2)$}.  
\par

\begin{figure}[hbtp]
\begin{center}
\leftskip-.1cm
{\includegraphics[height=3.5cm]{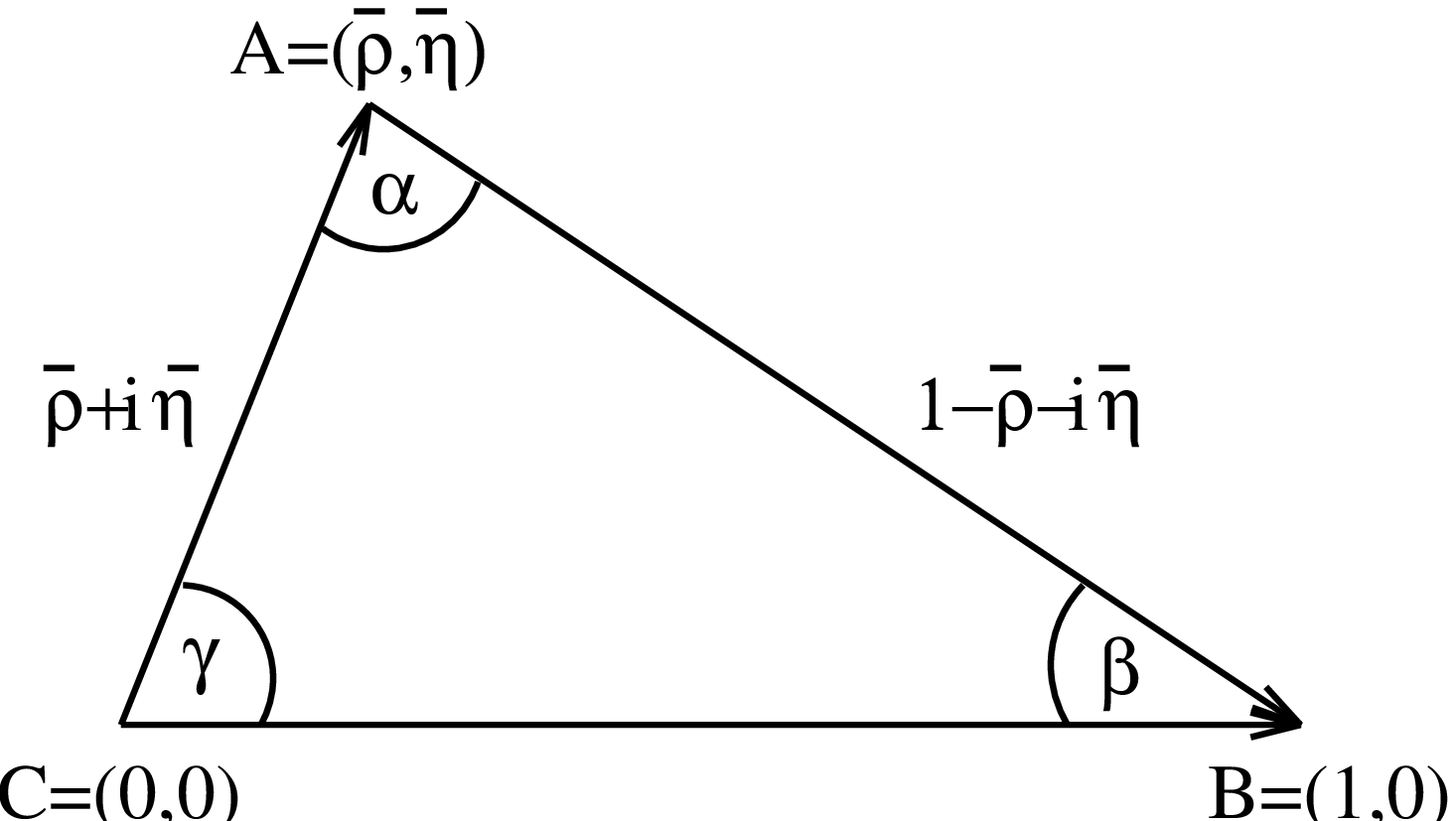}}
\vskip-6.5cm
\leftskip7cm
{\includegraphics[height=9.0cm]{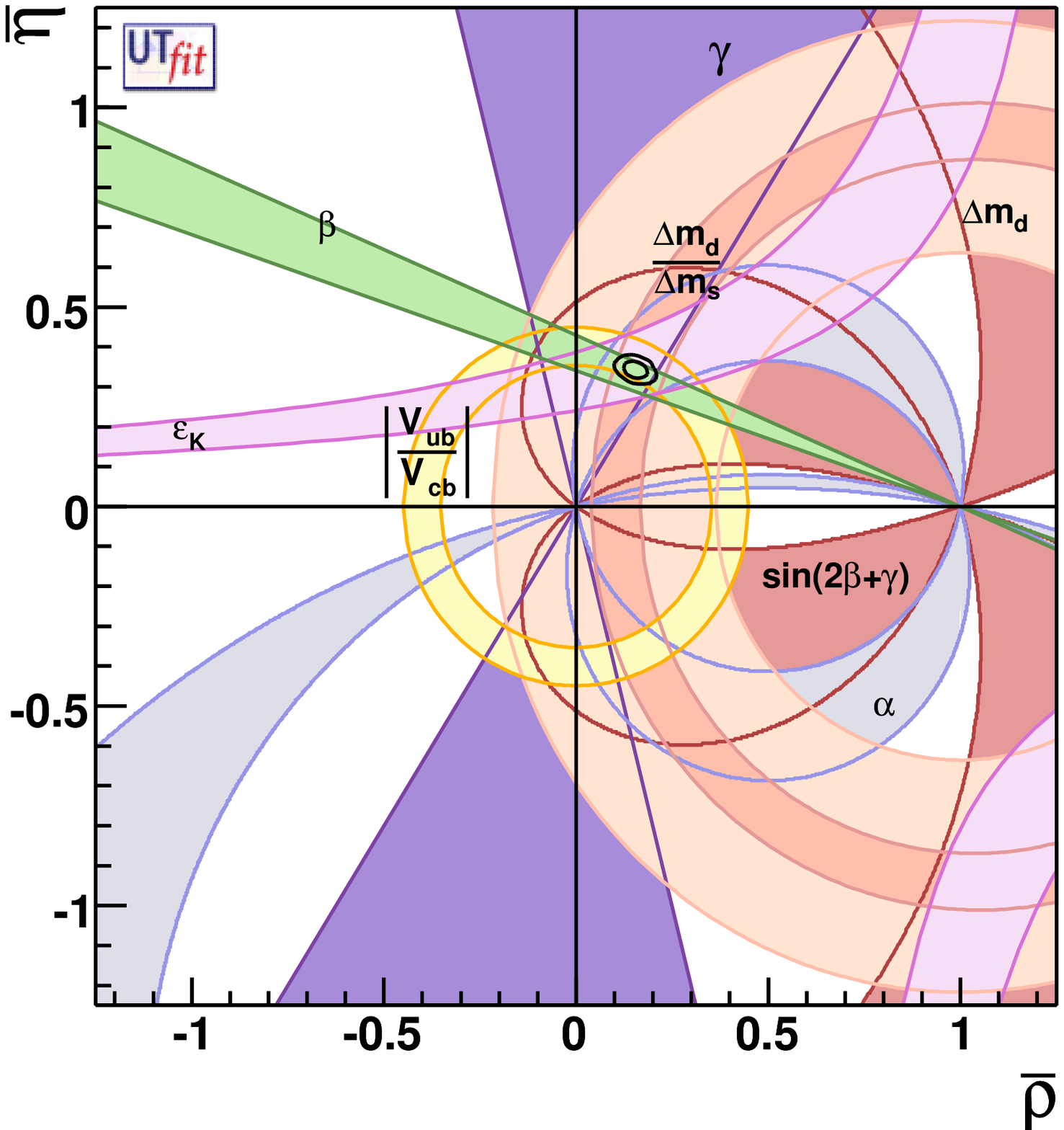}}
\caption{Left: Unitarity triangle in the $(\rhobar,\etabar)$ plane. 
$(1-\rhobar -i\etabar)$ corresponds to $\frac{V_{td}}{\lambda V_{cb}}$ and $(\rhobar + i\etabar)$ to 
$\frac{V^*_{ub}}{\lambda V_{cb}}$. Right: experimental constraints in the $(\rhobar,\etabar)$ plane. The closed 
contours show the most probable region at 68~\% and 95~\% CL. The individual constraints (given at 95 \% CL)  are shown 
by the coloured regions. }
\label{fig:triangle}
\end{center}
\end{figure}

The unitarity of the CKM matrix imposes a set of relations among its elements, including the
condition $V_{ud} V_{ub}^{*}+V_{cd} V_{cb}^{*}+V_{td} V_{tb}^{*}=0$ which defines a triangle in the complex plane, shown in Fig.~\ref{fig:triangle}. 
Many measurements can be conveniently displayed and compared as constraints on
sides and angles of this triangle. CP violation is proportional to the area of the unitarity triangle
and therefore it requires that all sides and angles be different from zero. The angle 
$\gamma \equiv \phi_{3} = arg(-V_{ud} V_{ub}^{*}/V_{cd} V_{cb}^{*})$
is the least precisely known~\cite{bib:Fitters}. 
For accurate amplitude predictions from the Standard Model, the four parameters of the
CKM mass mixing matrix, in particular $\rhobar$ and $\etabar$, must be measured precisely. Furthermore,
this must be done in a way that the extracted parameters are not affected by
possible new physics contributions, i.e. done only with the quantities related to tree diagrams.
One of the most promising ways is to combine the measurement of $|V_{ub}/V_{cb}|$ via semileptonic decays of the $B$ meson, which
depends on $\sqrt{ \rhobar^2 + \etabar^2}$, and CP violation in $B \rightarrow DK$ decays, which gives $ \arctan (\etabar / \rhobar)$ (the
angle $\gamma / \phi_3$  of the unitarity triangle. 

\par
Different approaches have been used to measure the angle  $\gamma$ (or $\phi_3$)
of the unitarity triangle. They exploit the interference between $b \rightarrow c$  and $b \rightarrow u $  transitions.
Up to now, this has been achieved using decays of type \BtoDzK\ and \BtoDzbK\ with subsequent decays into final states 
accessible to both charmed meson and anti-meson\footnote{In all this paper the mention of a decay will refer also to its charge-conjugate state, 
except explicitely stated.}. They are classified in three main types: 
\begin{description}
\item [The GLW method]~\cite{bib:GLW}: the \Dz\ meson decays into a CP final state
\item [The ADS method]~\cite{bib:ADS}: the \Dz\ meson is reconstructed into the $K \pi$ final state, 
for the $b \rightarrow c$ ({\it resp.} $b \rightarrow u$) transitions the \Dz\ decay mode will be the Cabibbo 
suppressed: $\Dz \rightarrow K^+ \pi^-$ ({\it resp.} Cabibbo allowed: $\Dz \rightarrow K^- \pi^+$). 
In this way the magnitude of the two 
interfering amplitudes will not be too different.   
\item [The GGSZ method]~\cite{bib:GGSZ}: the \Dz\ final state is $\KS \pi \pi$ which is accessible to both 
\Dz and \Dzb. This requires analysis of the \Dz\ Dalitz plot. This strategy can be seen as a mixture of the two previous
ones, depending on the position in the Dalitz plot. 
\end{description}

If one writes $A(\BmtoDzKm) = A_ce^{i\delta_c}$, $A(\BmtoDzKm) = A_ue^{i\delta_u - \gamma}$,  $A(\Dz \rightarrow f) = A_fe^{i\delta_f}$ and
$A(\Dzb \rightarrow f) = A_{\overline{f}}e^{i\delta_{{\overline f}}}$ ($f$ is a generic final state of the $D$ meson) then
it can be shown that :
\begin{equation}
\Gamma \left( \Bm \rightarrow f_{D} \Km \right) = A_{c}^{2} A_{\overline{f}}^{2} \left( \frac{A_{f}^{2}}{A_{\overline{f}}^{2}} + r_B^2 +
\frac{2r_B A_{f} }{A_{\overline{f}}} \Re \left( {e^{i(\delta_B+\delta_D-\gamma)}} \right) \right) 
\label{eq:GammaGeneral}
\end{equation}
where $r_B = A_u/A_c$, $\delta_B = \delta_u -\delta_c$ and $\delta_D = \delta_f - \delta_{{\overline f}}$. The rate for the charge conjugate process
is obtained from Eq.\ref{eq:GammaGeneral} replacing $\gamma$ by $-\gamma$. 
The first measurements have been done by the B factories (BaBar and BELLE) which still have the most precise measurements. 
However, in both cases, the data taking 
period has ended and the $\gamma / \phi_3$ angle is still not precisely measured. 
An important milestone has been achieved by the CDF experiment, operating at the Tevatron proton-antiproton collider, which has demonstrated that
GLW or ADS analyses (see section~\ref{sec:CDF}) are possible 
in an hadronic environment.
A summary of the experimental conditions and event yields in a representative decay mode is given in Tab.~\ref{tab:exp}. 
\begin{table}[htb]
\begin{center}
\begin{tabular}{|l|l|l|l|l|}
 \hline
Experiment & $\sigma_{\bbbar}$ & $\sigma_{\mathrm{inel}}/ \sigma_{\bbbar}$ & Integrated luminosity & $B^- \rightarrow (K^- \pi^+)_{D} K^-$\\
        &     &     &  & yield \\
\hline
BaBar 	   &   $\sim 1 $ nb 	   & $\sim 4 $ 				& 425 fb$^{-1}$ 	& $\sim 1940$ \\	
BELLE 	   &   $\sim 1 $ nb 	   & $\sim 4 $ 				& 700 fb$^{-1}$ 	& $\sim 3400$ \\	
CDF        &   $\sim 100 \mu$b	   & $\sim 1000$			& $\sim $ 10 fb$^{-1}$ full dataset&  $\sim 1500$ (5 fb$^{-1}$) \\	
LHCb       &   $\sim 290 \mu$b	   & $\sim 300$			& $\sim $ 1 fb$^{-1}$ (2011 expected)&  $\sim 440$ (0.035 fb$^{-1}$) \\	
\hline
 \end{tabular}
\caption{Summary of the various experimental conditions. 
For the Tevatron and LHCb the yields correspond to those of the analyses so far performed.
The Cabibbo-allowed decay $B^- \rightarrow (K^- \pi^+)_{D} K^-$ provides an estimate of the 
selection and trigger efficiencies for the various detectors.}
\label{tab:exp}
\end{center}
\end{table}
Note also that with the LHCb experiment it will also be possible to measure  $\gamma / \phi_3$  in a \BsToDsK\ time dependent analysis, 
thanks to the large \Bs\ sample, excellent proper time resulution and particle identification capabilities of the experiment.
These proceedings are organized as follow: 
after having shown the available results from CDF, which will allow me to present in more details the GLW and ADS 
methods, I will switch to LHCb results. Indeed, even if LHCb is not yet able to perform 
$\gamma / \phi_3$ measurements with the very little integrated lumi collected in 2010, some encouraging results are already available.  

\section{CDF results}\label{sec:CDF}
\subsection {GLW measurements}
In the method proposed by GLW the \Dz\ meson is reconstructed to CP eigenstate final states. Due to the high background level at
hadronic colliders the CP=-1 (e.g. \KS $\pi^0$) modes are not accessible and in fact only the CP=+1 $\Kp \Km$ \Dz\ final state 
has been used. Using a CP=+1 
final state for the $D$ meson leads to a simplified form for Eq.\ref{eq:GammaGeneral} using $\delta_D=0$ and 
$\frac{A_{f}}{A_{\overline{f}}}=1$: 
\begin{equation}
\Gamma \left( \Bm \rightarrow f_{CP+} \Km \right) = A_{c}^{2} A_{f_{CP+}}^{2} \left( 1 + r_B^2 + r_B \cos \left( \delta_B-\gamma \right) \right) 
\label{eq:GammaGLW}
\end{equation}
If CP=-1 final states could be reconstructed the formula would be similar except a - sign in front of the cosine term. 
From Eq.\ref{eq:GammaGLW} one can derive the asymmetries which are experimentally measured: 
\begin{equation}
A_{CP+} = \frac{\Gamma \left( \Bm \rightarrow f_{CP+} \Km \right) - \Gamma \left( \Bp \rightarrow f_{CP+} \Kp \right)  }{\Gamma \left( \Bm \rightarrow f_{CP+} \Km \right) + \Gamma \left( \Bp \rightarrow f_{CP+} \Kp \right) } = \frac{2r_B\sin \delta_B \sin \gamma}{1+r_B^2+2r_B\cos \delta_B \cos \gamma} 
\label{eq:GammaGLW_A}
\end{equation}
\begin{equation}
R_{CP+} = \frac{\Gamma \left( \Bm \rightarrow f_{CP+} \Km \right) + \Gamma \left( \Bp \rightarrow f_{CP+} \Kp \right)  }{ \Gamma \left( \Bm \rightarrow \Dz \Km \right)+\Gamma \left( \Bp \rightarrow \Dzb \Kp \right)  } = 1+r_B^2+2r_B\cos \delta_B \cos \gamma \\
\label{eq:GammaGLW_R}
\end{equation}
in case of a CP=-1 $D$ meson final state one would have: 
\begin{equation}
A_{CP-} =  \frac{-2r_B\sin \delta_B \sin \gamma}{1+r_B^2-2r_B\cos \delta_B \cos \gamma} \\
R_{CP-} =  1+r_B^2-2r_B\cos \delta_B \cos \gamma \\
\end{equation}
It should be noted that only 3 of these 4 ratios are independent while there are 3 unknowns. In addition, due to the formulae themselves 
there is an 8-fold ambiguity. In summary, the GLW measurements have a poor constraining power on $\gamma / \phi_3$
when they are considered alone. However, they generally improve the knowledge of $r_B$, $\gamma / \phi_3$ and 
$\delta_B$ when combined with other methods (for example the ADS method).
The CDF collaboration has performed an analysis~\cite{bib:CDF_GLW} on a sample collected by the upgraded 
CDF detector corresponding to an integrated 
luminosity of $\sim 1$ fb$^{-1}$. Both $\Bm \rightarrow \Dz \Km $ and $\Bm \rightarrow \Dz \pim $ decay modes are reconstructed 
and the 
$\Dz$ in the $\Km \pip$ flavour final state and the two CP=+1 final states $\Kp \Km$ and $\pip \pim$. 
The $\Bm \rightarrow \Dz \pim $ modes are much more abundant than the $\Bm \rightarrow \Dz \Km $ and provide useful constraints
 in the 
fit. An unbinned maximum likelihood fit which combines invariant $B$ mass assuming that the bachelor track is a pion, momenta,
 and 
Particle IDentification (PID) information obtained from dE/dx measurement for all three modes of interest is performed simultaneously. 
The measured yields are given in Tab.~\ref{tab:CDF_GLW} and the final ratios are: 
\begin{eqnarray}
\frac{\Gamma \left( \Bm \rightarrow \Dz \Km \right) + \Gamma \left( \Bp \rightarrow \Dzb \Kp \right)}{\Gamma \left( \Bm \rightarrow \Dz \pim \right) + \Gamma \left( \Bp \rightarrow \Dzb \pip \right)  }
= 0.0745 \pm 0.0043 \mathrm {(stat.)} \pm 0.0045 \mathrm{(syst.)} \\
R_{CP+} = 
\frac{\Gamma \left( \Bm \rightarrow f_{CP+} \Km \right) + \Gamma \left( \Bp \rightarrow f_{CP+} \Kp \right)  }
{ \Gamma \left( \Bm \rightarrow \Dz \Km \right) }
= 1.30 \pm 0.24 \mathrm {(stat.)} \pm 0.12 \mathrm{(syst.)} \\
A_{CP+} = \frac{\Gamma \left( \Bm \rightarrow f_{CP+} \Km \right) - \Gamma \left( \Bp \rightarrow f_{CP+} \Kp \right)  }{\Gamma \left( \Bm \rightarrow f_{CP+} \Km \right) + \Gamma \left( \Bp \rightarrow f_{CP+} \Kp \right) }
= 0.39 \pm 0.17 \mathrm {(stat.)} \pm 0.04 \mathrm{(syst.)} \\
\end{eqnarray}
\begin{table}[htb]
\begin{center}
\begin{tabular}{|l|l|l|}
 \hline
                        & $ \Bp \rightarrow D \Kp $  & $\Bm \rightarrow D \Km  $  \\ \hline
 $D \rightarrow K \pi$  & $250\pm 26$                       &   $266\pm 27$                       \\    
 $D \rightarrow K K$    & $22\pm 8$                         &   $49\pm 11$                       \\    
 $D \rightarrow \pi \pi $ & $6\pm 6$                       &   $14\pm 6$                       \\    
\hline
 \end{tabular}
\caption{Summary of the measured yields in the CDF GLW analysis.}
\label{tab:CDF_GLW}
\end{center}
\end{table}

For all these three ratios, one of the main source of systematical uncertainty is the PID. Indeed the kaon pion separation 
is only at 1.5 standard deviation for a 2 GeV momentum track. Another source of systematical uncertainty is the handling of 
the combinatorial background 
and this will improve with a larger data sample. The CDF collaboration has now recorded more than 10 times more data ; 
this should allow them to perform
 a more accurate measurement in the future. 

\subsection {ADS measurements}
\begin{figure}[htbp]
    \begin{center}
        \begin{minipage}[t]{0.49\textwidth}
            \begin{center}
              \includegraphics[width=7.5cm, height = 4.5 cm]{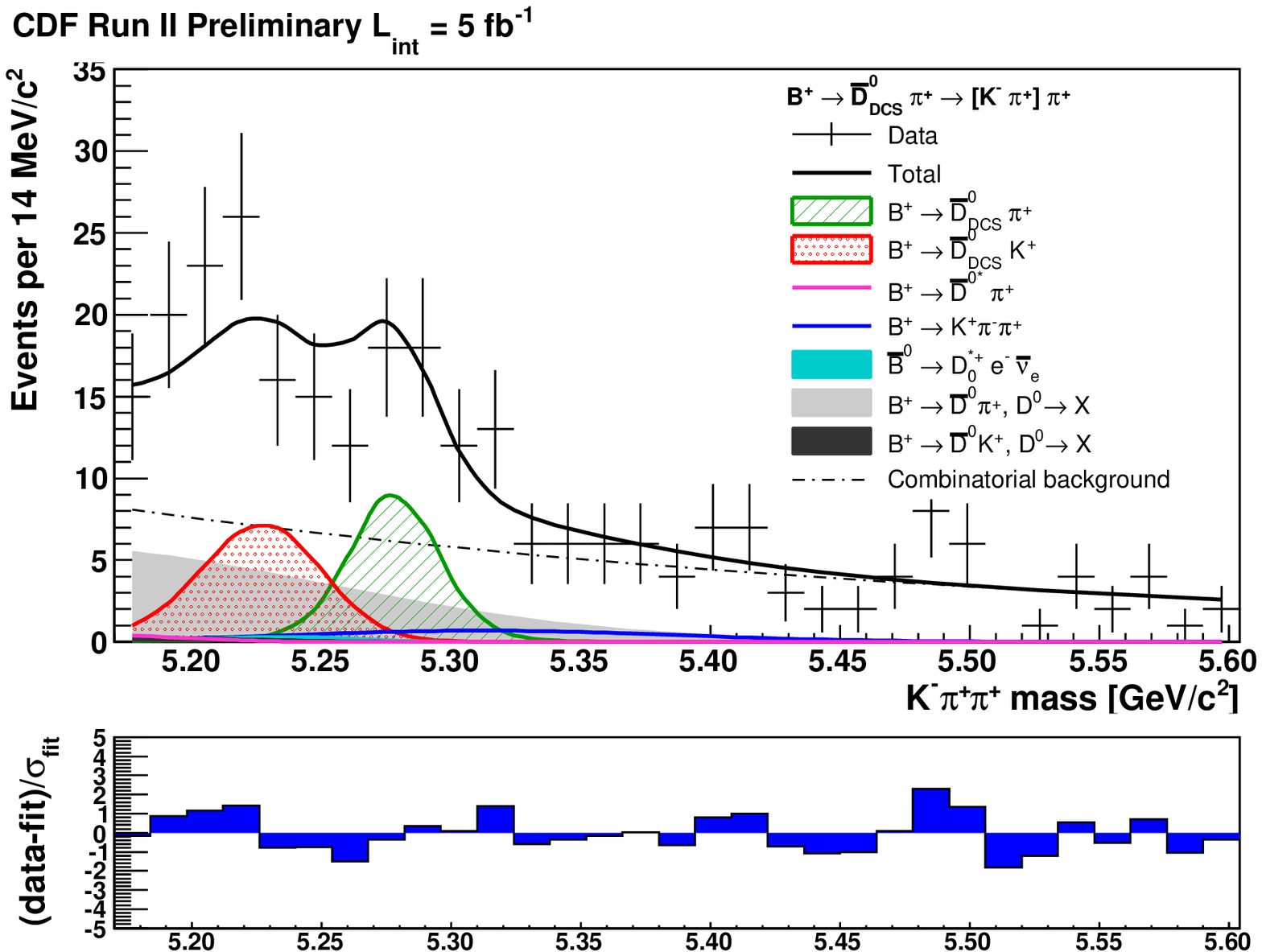}
            \end{center}
        \end{minipage}
	 \begin{minipage}[t]{0.05\textwidth}
	 \end{minipage}	
        \begin{minipage}[t]{0.49\textwidth}
            \begin{center}
            	\includegraphics[width=7.5cm, height = 4.5 cm]{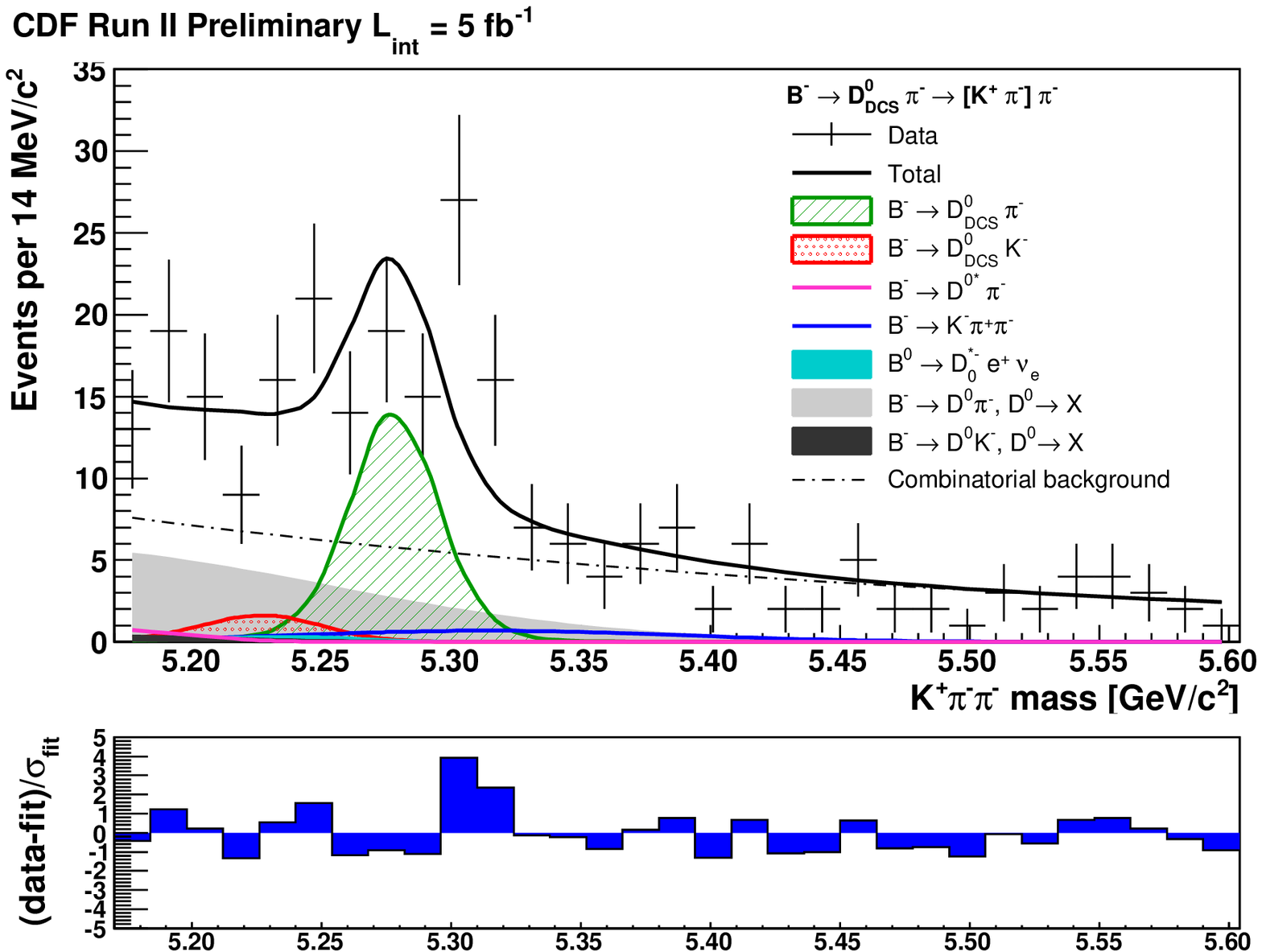}
            \end{center}
        \end{minipage}
    \end{center}
\caption{Fit projection onto the \Dz $\pi$ variable for the doubly Cabibbo-suppressed mode, positive charges (left) and negative charges (right).}
\label{fig:ads}
\end{figure}

The ADS measurement performed at the Tevatron collider by the CDF experiment~\cite{bib:CDF_ADS} uses about 5~fb$^{-1}$. The \Dz\ meson is reconstructed using both the 
Cabibbo Favoured (CF) $\Dz \rightarrow K^- \pi^+$ ($D_{CF}$) and the Cabibbo Suppressed (CS) $\Dz \rightarrow \Kp \pim$ ($D_{CS}$) modes. 
The analysis strategy is similar to the one developed for the 
GLW measurement: not only the $ \Bm \rightarrow D \Km  $ is reconstructed but also the $ \Bm \rightarrow D \pim  $ decay mode. 
The cuts optimization is focused on finding an evidence of the  $ \Bm \rightarrow D_{CS} \pim  $ mode. 
Since the $ \Bm \rightarrow D_{CF} \pim  $  mode has the same topology of the CS one, the cuts optimization is performed on the 
CF mode.
The optimized cuts are those related to the flight of the $D$ and $B$ meson, the isolation of the $B$ meson and PID. The fit 
is an unbinned maximum 
likelihood fit simultaneously performed on the CF and DCS modes and spliting the samples according to the $B$ charge. It uses 
two discriminating 
variables:  the $B$ mass (asumming a pion mass for the bachelor track) and the PID information for the bachelor track. 
The projection on the $B$ mass for (CS) decays is shown in Figs.~\ref{fig:ads}. 
The fitted yields can be translated into asymmetries both for the $\Bm \rightarrow \Dz \Km $ and $\Bm \rightarrow \Dz \pim $ 
decay modes.
\begin{eqnarray*}{}
R_{ADS}(\pi) &=& \frac{\Gamma \left( \Bm \rightarrow D_{CS} \pim \right) + \Gamma \left( \Bp \rightarrow D_{CS} \pip \right)}
{\Gamma \left( \Bm \rightarrow D_{CF} \pim \right) + \Gamma \left( \Bp \rightarrow D_{CF} \pip \right)} 
= \left( 4.1 \pm 0.8 {\mathrm (stat.)} \pm 0.4 {\mathrm (syst.)} \right) . 10^{-3} \\
R_{ADS}(K) &=& \frac{\Gamma \left( \Bm \rightarrow D_{CS} \Km \right) + \Gamma \left( \Bp \rightarrow D_{CS} \Kp \right)}
{\Gamma \left( \Bm \rightarrow D_{CF} \Km \right) + \Gamma \left( \Bp \rightarrow D_{CF} \Kp \right)} 
= \left( 22.5 \pm 8.4 {\mathrm (stat.)} \pm 7.9 {\mathrm (syst.)} \right) . 10^{-3} \\
A_{ADS}(\pi) &=& \frac{\Gamma \left( \Bm \rightarrow D_{CS} \pim \right) - \Gamma \left( \Bp \rightarrow D_{CS} \pip \right)}
{\Gamma \left( \Bm \rightarrow D_{CS} \pim \right) + \Gamma \left( \Bp \rightarrow D_{CS} \pip \right)} 
=  0.22 \pm 0.18 {\mathrm (stat.)} \pm 0.06 {\mathrm (syst.)}  \\
A_{ADS}(K) &=& \frac{\Gamma \left( \Bm \rightarrow D_{CS} \Km \right) - \Gamma \left( \Bp \rightarrow D_{CS} \Kp \right)}
{\Gamma \left( \Bm \rightarrow D_{CS} \Km \right) + \Gamma \left( \Bp \rightarrow D_{CS} \Kp \right)} 
=  -0.63 \pm 0.40 {\mathrm (stat.)} \pm 0.23 {\mathrm (syst.)}  \\
\end{eqnarray*}
These results are in agreement with B-factories measurements~\cite{bib:hfag} and the precision is similar to the one obtained
by the BaBar collaboration.

\section{LHCb results}
From the two measurements described above, the CDF collaboration has proven that $B$ physics with pure hadronic decays is also 
accessible to hadron colliders experiments. At the time of FPCP conference the LHCb collboration had only presented results 
based on 2010 data ($\sim 36$ pb$^{-1}$)\footnote{More than 10 times more data is already on tape at the time of
 writing these proceedings}. I will present some measurements made using decays similar to the ones used for 
$\gamma / \phi_3$.

\subsection{LHCb performances} 
Despite the fact that the LHCb detector has been taking data for only one year, 
the experimental attributes critical for the $\gamma / \phi_3$ analysis are already demonstrating a performance 
adequate for the measurement.  These attributes include the hadronic trigger, the 
displaced vertices reconstruction and the PID for charged hadrons. 
LHCb operates a two-level trigger system: a hardware trigger (L0) and a software implemented High Level Trigger (HLT). 
The L0 trigger provided by the ECAL, MUON, and HCAL information reduces the visible interaction rate from 10 MHz to 1 MHz ; 
the HLT reduces the trigger rate further to 2 kHz. During the 2010 data taking period, several trigger configurations were used 
both for the L0 and the HLT in order to cope with the varying beam conditions. For the hadronic trigger the threshold was set 
to 3.6 GeV of transverse momentum for most of the data. The HLT confirms the L0 information using tracking data and adds 
some transverse momentum and impact parameter requirements. The L0 efficiency was of the order of 50\% for our channels while 
the HLT one was more of the order of 80\%. Note however that triggering is also possible on tracks from the other $B$ decay present in the event. 
The impact parameter resolution for tracks with large transverse momentum 
is of the order of 15 $\mu m$ and the  resolution on the primary vertex reconstruction is of the order of 15 
 $\mu m$ in the transverse plane and 75  $\mu m$ along the beam axis. At LHC, due to the large center of mass energy the $B$ 
flies on average 1 cm before decaying. Finally, the PID performances as a function of the track momentum, are measured on data
 using the K and $\pi$ from the \Dz decay originating from the decay of a \Dstar\ with kinematical properties similar to 
our signal. On average, the 
kaon efficiency is of the order of 95\% for a pion contamination of 7\%. 

\subsection{First signal yields for  $\Bm \rightarrow D^0 \Km$ decays }
Using 2010 data ($\sim 36$ pb$^{-1}$), the LHCb experiment has been able to extract clean samples of both 
$\Bm \rightarrow \Dz \Km $ and $\Bm \rightarrow \Dz \pim $ decay modes.
 The good PID performances and impact parameter resolution are of critical importance for the signal 
extraction from the large hadronic background. The rescaled yields per fb$^{-1}$ are summarized in Tab~\ref{tab:Yields_Dh} for LHCb and CDF. 
From these numbers, it is clear that the results expected for the summer conferences from LHCb should be competitive with those from CDF presented
at this conference. 
\begin{table}[htb]
\begin{center}
\begin{tabular}{|l|l|l|}
 \hline
                        & CDF  & LHCb  \\ \hline
 $\B^{\pm} \rightarrow (K \pi)_D \pi^{\pm} $ &    $\sim$ 3.5k                        &   $\sim$ 223k                       \\    
 $\B^{\pm} \rightarrow (K \pi)_D K^{\pm} $   &    $\sim$ 0.3k                        &   $\sim$ 12.6k                       \\    
 $\B^{\pm} \rightarrow (K K)_D \pi^{\pm} $   &    $\sim$ 0.8k                        &   $\sim$ 28.6k                       \\    
\hline
 \end{tabular}
\caption{Comparison of the measured yields rescaled to an integrated luminosity of 1 fb$^{-1}$. The choice is made to 
give numbers only for the modes studied  in the 2010 data set by LHCb.}
\label{tab:Yields_Dh}
\end{center}
\end{table}

\subsection{First observation of the decay $\overline{B}_s \rightarrow D^0 K^{*0}$ }
\begin{figure}[htbp]
    \begin{center}
              \includegraphics[width=9.5cm, height = 7.5 cm]{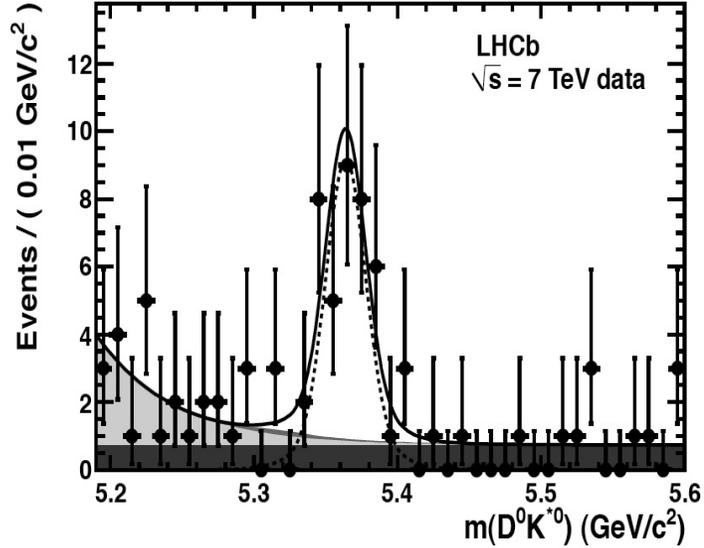}
            \end{center}
\caption{Invariant mass distribution for the \BsbartoDKstar\ decay mode with the fit superimposed. 
The black points correspond 
to the data and the fit result is represented as a solid line. The signal is fitted with a double Gaussian (dashed line), 
the partially reconstructed with an exponential function (light grey area), the combinatorial background with a flat distribution 
(dark grey area) and the cross-feed from \BdbartoDRho\ (intermediate grey area). }
\label{fig:Bs_DKstar}
\end{figure}

Another promising way to measure $\gamma / \phi_3$ is to use the $B_d \rightarrow D K^{*0}$ decay mode where $D$ represents an admixture
of $D^0$ and $\bar D^0$ mesons. 
Although the channel involves the decay of a neutral B meson, the final state is self-tagging so that a time-dependent 
analysis is not required. Since both the CKM favoured and suppressed decays are also colour suppressed, the branching 
fractions are smaller than the equivalent charged $B^{\pm} \rightarrow D K^{\pm}$ decays, but exhibit an
enhanced interference.
The Cabibbo-allowed $B_s$ decays, $\overline{B}_s \rightarrow D^0 K^{*0}$ and $\overline{B}_s \rightarrow D^{*0} K^{*0}$, potentially cause
a significant background to the Cabibbo-suppressed $B_d \rightarrow D^0 K^{*0}$ decay. Moreover, the
expected size of this background is unknown, since the $\overline{B}_s \rightarrow D^{(*)0} K^{*0}$ decay has not yet
been observed. In addition, a measurement of the branching fraction of $\overline{B}_s \rightarrow D^0 K^{*0}$ is
of interest as a probe of SU(3) breaking in colour suppressed $B \rightarrow D^0 V^{0}$ decays. The
detailed study of $\overline{B}_s \rightarrow D^0 K^{*0}$  is thus an important and interesting milestone towards the
measurement of  $\gamma / \phi_3$.
The strategy of the analysis is to measure a ratio of branching fractions in which most of the potentially large systematic 
uncertainties cancel. The decay $B_d \rightarrow D^0 \rho^0$ for which about two to eight times more events are expected than for the 
$\overline{B}_s \rightarrow D^0 K^{*0}$ decay, is used as the normalisation channel. In both decay channels, the $D^0$ is reconstructed 
in the Cabibbo-favoured decay mode $D^0 \rightarrow \Km \pi^+$; the contribution from the doubly Cabibbo-suppressed decay is negligible. 
The main systematic uncertainties arise from the different particle identification requirements and the pollution of the 
$B_d \rightarrow D^0 \rho^0$ peak by $B_d \rightarrow D^0 \pi^+ \pi^-$ decays where the $\pi^+ \pi^-$ pairs do not originate from a 
$\rho^0$ resonance. In addition, the normalization of the $B_s$ decay to a $B_d$ decay suffers from a systematic uncertainty of the order 
of about 13 \% due to the ratio of the fragmentation fractions $f_d/f_s = 3.71 \pm 0.47$. The $B_s \rightarrow D^0 K^{*0}$ signal peak obtained on 
2010 data is shown on Fig.~\ref{fig:Bs_DKstar}. 

A clear signal of $35.5 \pm 6.9$ \BsbartoDKstar\ events is observed for the first time.   
The probability of the background fluctuating to form the \Bs\ signal corresponds to 
approximately nine standard deviations,
as determined from the change in twice the natural logarithm of the likelihood of the fit without signal. Although this significance includes 
the statistical uncertainty only, 
the conclusion is unchanged if the small sources of systematic error that affect the yields are included. 
The branching ratio for this decay is measured relative to that for
\BdbartoDRho to be
%\[
$$
\frac{\BR\left(\BsbartoDKstar\right)}{\BR\left(\BdbartoDRho\right)} = 1.39 \pm 0.31 \pm 0.17 \pm 0.18
$$
%\] 
where the first uncertainty 
is statistical, the second systematic and the third is due to the hadronisation fraction ($f_d/f_s$).

\subsection{First observation of the decay $B_d \rightarrow D^0 K  \pi \pi$ }
\begin{figure}[htbp]
    \begin{center}
              \includegraphics[width=9.5cm, height = 7.5 cm]{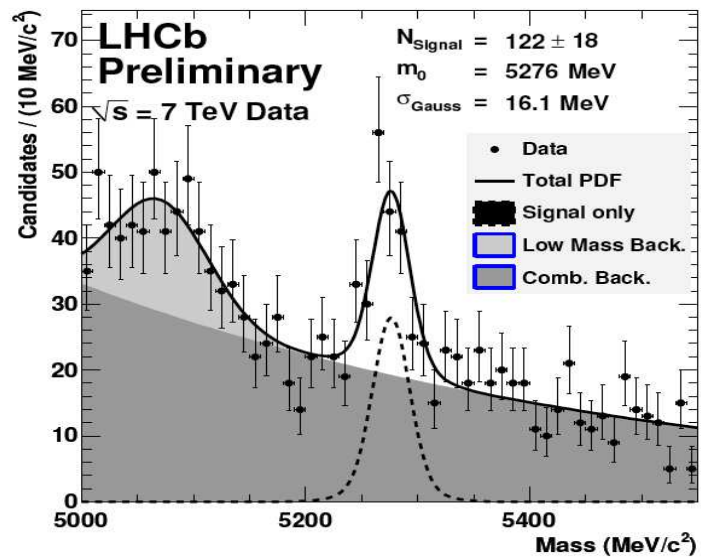}
            \end{center}
\caption{
Invariant mass distribution for the $B_d \rightarrow D^0 K \pi \pi$ mode obtained on 2010 data. 
The black points correspond 
to the data and the fit result is represented as a solid line. The background is due to partially reconstructed 
events (the shape in light grey is taken from Monte-Carlo)  and to combinatorial background (drak grey) modelled with a second order polynomial.}
\label{fig:LHCb_DK2Pi}
\end{figure}

In order to better understand $B$ decays and eventually add new modes for the $\gamma / \phi_3$ measurement, the LHCb collaboration has studied 
various $B \rightarrow D 3h$ decays \cite{bib:LHCb-CONF-2011-024}. 
Among other signals, a clear peak of $122 \pm 18$ events is obtained for the  $B_d \rightarrow D^0 K \pi \pi$ mode leading to 
a measurement of ratio of branching ratios : 

%\[
$$
\frac{\BR\left(B_d \rightarrow D^0 K \pi \pi  \right)}{\BR\left(B_d \rightarrow D^0 \pi \pi \pi\right)} = 9.6 \pm 1.5 \pm 0.8
$$
%\] 
The signal plot is shown on Fig.~\ref{fig:LHCb_DK2Pi}. 
 
\subsection{Time dependent measurements} 
Information about $\gamma / \phi_3$ can also be obtained by other means: time dependent measurements using the $B_s \rightarrow D_s K$ decay mode 
or 2-body charmless $B$ events. The data collected in 2010 do not allow for such measurements. However the $B_s$ mixing frequency 
has been measured as well its lifetime using the charmless decay $B_s \rightarrow  K K$. 
More details are given in ~\cite{bib:LHCb-CONF-2011-005} and ~\cite{bib:LHCb-CONF-2011-018} and in Marta Calvi's contribution to these proceedings.

\Acknowledgements
I would like to thank the FPCP organizers for this very nice conference with many interesting discussions, my LHCb colleagues as well as the CDF 
collaboration for providing the material discussed here. I am grateful to Aur\'elien Martens and Guy Wilkinson for their careful 
reading of these proceedings.

\end{document}